\documentclass[modern]{aastex61}
\usepackage{graphicx}
\usepackage{times}
\usepackage{float}
\usepackage{amsmath}
\usepackage{epsfig}
\usepackage{subfigure}
\usepackage{natbib}
\usepackage[normalem]{ulem}
\usepackage{censor}
\usepackage{gensymb}
\newlength\nextcharwidth
\makeatletter
\renewcommand\@cenword[1]{%
  \setlength{\nextcharwidth}{\widthof{#1}}%
  \censorrule{\nextcharwidth}%
  \kern -\nextcharwidth%
  #1}
\makeatother

\renewcommand*{\thesection}{\Roman{section}}
\renewcommand*{\thesubsection}{\thesection.\Alph{subsection}}

\shorttitle{GCM Climate Instability}
\shortauthors{Paradise, Menou}

\begin{document}
\title{GCM Simulations of Unstable Climates in the Habitable Zone}
\author{Adiv Paradise}
\affiliation{Department of Astronomy and Astrophysics, University of Toronto, St. George}

\author{Kristen Menou}
\affiliation{Department of Astronomy and Astrophysics, University of Toronto, St. George}
\affiliation{Centre for Planetary Sciences, Department of Physical and Environmental Sciences, University of Toronto, Scarborough}

\correspondingauthor{Adiv Paradise}
\email{paradise@astro.utoronto.ca}




\submitjournal{The Astrophysical Journal}


\begin{abstract}

It has recently been proposed that Earth-like planets in the outer regions of the habitable zone experience unstable climates, repeatedly cycling between glaciated and deglaciated climatic states \citep{menou2015}. While this result has been confirmed and also extended to explain early Mars climate records \citep{haqqcycles,earlymars}, all existing work relies on highly idealized low-dimensional climate models. Here, we confirm that the phenomenology of climate cycles remains in 3D Earth climate models with considerably more degrees of freedom. To circumvent the computational barrier of integrating climate on Gyr timescales, we implement a hybrid 0D-3D integrator which uses a general circulation model (GCM) as a short relaxation step along a long evolutionary climate sequence. We find that GCM climate cycles are qualitatively consistent with reported low-dimensional results. This establishes on a firmer ground the notion that outer habitable zone planets may be preferentially found in transiently glaciated states.

\end{abstract}

\section{Introduction}

Recent estimates of the habitable zone for Earth-like planets with modern Earth's rotation rate place an inner edge at 0.99 AU and an outer edge at 1.7 AU \citep{kopparapu13}. However, these estimates come from 1D climate models which only consider equilibria on short timescales. That short-term equilibrium comes from an energy balance requirement---the heat released into space must balance the net solar radiation (insolation) received from the sun \cite{north1981}. We refer to this as thermal equilibrium. Planets can also demonstrate stability or instability on longer timescales. Changes in glaciation or greenhouse gases over time can alter the planet's albedo and outgoing thermal emission, affecting the thermal equilibrium temperature \citep{north1981,pierrehumbert05}. It may be that a more realistic habitable zone that takes into account long-term stability will be significantly more constrained. The emergence of complex life depends as much on the long-term stability of a habitable climate as it does on its immediate habitability---an informed search for Earth analogs thus depends strongly on understanding climate variation throughout the habitable zone.

A planet's climate stability can be tied to the carbon-silicate cycle. Atmospheric carbon dioxide dissolves into rainwater as carbonic acid, is delivered to the surface, reacts with the rock, and is eventually carried to the oceans, where it can be recycled into the mantle via subduction \citep{Berner2004,pierrehumbertbook}. On planets containing vascular land plants, the direct dependence of weathering on the CO$_2$ partial pressure (hereafter pCO$_2$) is weakened by plant activity affecting carbon levels in the soil \citep{Berner2004,pierrehumbertbook,schartz89}. However, on planets lacking land plants, the carbon cycle is more strongly-coupled and acts as a negative feedback on the planet's surface temperature---increased temperatures or pCO$_2$ surpluses lead to increased weathering and CO$_2$ sequestration, while a deficit reduces weathering, allowing volcanic outgassing to replenish CO$_2$ \citep{pierrehumbertbook,walker81,schartz89,wk97,kump00}. The climate on such a planet tends toward an equilibrium point. At low insolation, the equilibrium point may be cold enough for total glaciation \citep{menou2015,pierrehumbert05,hoffman98}, called a ``snowball" state. \citet{Budyko1969} showed that such states can exist at the same insolations and CO$_2$ partial pressures as temperate climates, and therefore require either high insolation or significant greenhouse gases for deglaciation. In this scenario of a snowball state triggered by weathering, weathering ceases in the absence of liquid water, implying that no weathering equilibrium can be reached. Volcanic outgassing continues, and at a certain point pCO$_2$ will be high enough for the ice to melt, resulting briefly in warm wet conditions, followed by rapid weathering and cooling. Thus, on geological timescales the climate may cycle between long snowball events and brief warm wet periods \citep{menou2015}. Evidence of glacial deposits at then-tropical latitudes on Earth has been linked to this hypothesis \citep{Tajika2007}, and previous studies \citep[e.g.][]{pierrehumbert05,lucarini2010,Voigt2011} have shown that such climates are not ruled out by 3D climate models. We do not however aim to determine whether such episodes on Proterozoic Earth were triggered by weathering and later deglaciated by this process as opposed to other external triggers; there are sufficient uncertainties in both our model and Earth's conditions at the time that more sophisticated tools would be necessary.

\citet{menou2015} used a zero-dimensional reduction of 1D models to approximate energy-balanced climates of Earth-like planets at varying orbital distances throughout the habitable zone. He found that climate cycles, wherein the planet oscillates between brief warm periods and long frozen periods, should dominate much of the outer habitable zone. However, energy-balance models (EBMs) necessarily miss or must approximate inherently 3D phenomena and feedbacks, and so are of limited utility for predicting new climate phenomena. In the case of climate cycles, glaciation and deglaciation are dependent on factors such as sea ice distribution, precipitation patterns, heat transport by wind streams, local heating and cooling by clouds, seasonal melting, etc. Stronger or weaker regional temperature gradients can mean that the onset of snowball glaciation happens at global average surface temperatures that deviate significantly from 273.15 K. Since weathering is dependent on surface temperature, cooling or warming of a region due to changes in precipitation, cloud activity, and air streams as the climate changes can have a significant on weathering. It is therefore necessary to verify the existence of climate cycles using more sophisticated 3D climate models that can capture a diversity of climate feedbacks.
 
\section{Methods}
\subsection{Simulating Long Timescales}
\label{sec:longt}
Since climate cycles, as observed in 1D models by \citet{menou2015} and \citet{haqqcycles}, have periods of at minimum millions of years, the primary obstacle to simulating them in 3D models is the computational cost of millions of years or more of climate evolution in existing 3D models. PlaSim, the GCM used in this study, takes on the order of minutes of wall-time to compute one simulated year. Computing millions of years directly is thus unfeasible. A more detailed discussion of the computational cost of direct computation vs. our method is presented in \autoref{appendix1}. An approach is needed that allows a model to skip through millions of years of evolution, only computing a couple years at certain key points. 

We achieve this by noting that the climate exhibits different types of equilibria, which operate on different timescales. Thermal equilibrium, when the planet exhibits a relatively unchanging seasonally-averaged global temperature, is achieved on the order of years to decades following a perturbation. Weathering equilibrium, on the other hand, when global outgassing and weathering rates cancel each other out and the global pCO$_2$ level is relatively constant, is achieved (if at all) on much longer geological timescales of millions of years. This latter timescale is many orders of magnitude longer than the thermal timescale, which means that weathering effects can typically be disregarded on thermal timescales, and vice versa. The CO$_2$ partial pressure is a global quantity, which means that only global, zero-dimensional terms are needed to compute its evolution. Thermal equilibrium, however, must be computed using a full 3D treatment to capture various mechanisms such as sea ice feedback, wind circulation, cloud formation, etc. The global weathering rate also depends on 3D features such as landmasses and ice caps. Therefore, we construct a hybrid 3D-0D model in which a full 3D model computes the evolution of a new thermal equilibrium following a 0D perturbation in pCO$_2$. We compute the size and duration of that perturbation by computing the global weathering rate from the results of the 3D model. This means that millions of years of evolution of slow change can instead be approximated by allowing the CO$_2$ partial pressure to change by an equivalent amount, and then allowing the climate to relax back to thermal equilibrium. We neglect the intermediate timescale on which the atmospheric and oceanic carbon reservoirs equilibriate, assuming that partitioning is instantaneous. This reduces the computational cost, at the expense of model accuracy. In this way, we treat the GCM not as the primary driver of model evolution, but as a relaxation step in a simpler 0D model that controls the long-term model evolution.

This is very similar to the timescale-coupling approach used in \citet{Donnadieu2006} and \citet{Lehir2009} in their studies of the Pangea breakup and the Snowball Earth episode's aftermath, respectively. However, their approach involved sampling the expected evolutionary track and then interpolating between points using lookup tables to compute the time-dependence of the climate's evolution. Because our aim is to determine the very nature of the evolutionary track and to establish its time-dependence, we take a slightly different approach by extrapolating forward in time without \textit{a priori} knowledge of what the future climate states will look like, computing timescales using evaluated rather than interpolated weathering fluxes to determine timescales. 

For the 3D relaxation step, we use the Planet Simulator (PlaSim) developed at the Universit\"at Hamburg, a 3D Earth-system general circulation model (GCM) developed for fast and easy study of Earthlike climate systems. PlaSim solves primitive equations for vorticity, temperature, divergence, and pressure using the spectral transform method. It includes a 50-meter mixed-layer slab ocean, thermodynamic sea ice, a bucket model of soil hydrology, moist and dry convection, interactive clouds, large-scale precipitation, and a parameterized longwave and shortwave radiation model \citep{Fraedrich2005}. It uses a latitude-longitude grid with an atmosphere discretized into multiple vertical levels. For a more complete description of the model, readers are referred to the PlaSim Reference Manual \citep{plasimdocs}. PlaSim has been used before to study snowball climates and to map climate parameter spaces of Earth-like planets \citep{lucarini2010,Boschi2013,Lucarini2013}. PlaSim succeeds at reproducing the snowball hysteresis described in \citet{Budyko1969} and \citet{pierrehumbert05} \citep{lucarini2010}, and yields results consistent with the findings in \citet{pierrehumbert05} that snowball climates have greater seasonal variability, particularly in the troposphere lapse rate and tropopause height, and demonstrate on average weaker lapse rates and a lower tropopause. In PlaSim, snowball deglaciation is associated with the onset of melting sea ice in the tropics. We note that \citet{Abbot2010} has called this deglaciation mechanism into question as a possible artifact of simplistic sea ice schemes.

PlaSim however is a model of intermediate complexity, and while it qualitatively reproduces the snowball climates described by \citet{pierrehumbert05} and others, it necessarily makes approximations, parameterizations, and omissions as a tradeoff for a large advantage in speed. The aim of this paper therefore is not to thoroughly investigate the dynamics of a snowball climate or determine precise quantitative limits on glaciation and deglaciation, as slower, more sophisticated models would be better-suited to that task, but rather to take advantage of PlaSim's speed and simulate long timescales across wide swaths of parameter space, to verify whether or not climate cycles persist when investigated with models of increasing complexity and to identify overall trends in the habitable zone parameter space.

Our weathering implementation is based on a parameterization of Earth's silicate weathering in the absence of vascular land plants. We use the same form used in \citet{menou2015}, which is a form of the parameterization introduced in \citet{Berner1994} and expanded on in \citet{Berner2001}. That parameterization is based on observations of cation concentrations in river runoff, laboratory measurements, and seafloor drill cores. Our functional weathering form is thus 
\begin{equation}\label{eq:carbon}
\frac{W}{W_\oplus} = \left(\frac{\text{pCO}_2}{\text{pCO}_{2,\oplus}}\right)^{\beta}e^{[k_\text{act}(T_\text{s}-288)]}[1+k_\text{run}(T_\text{s}-288)]^{0.65},
\end{equation}
where $T_s$ is the surface temperature, pCO$_{2,\oplus}$ is the pre-industrial pCO$_2$ of 330 $\mu$bar, and $W_\oplus$ is the weathering rate at $T_\text{s}$ = 288 K. $k_\text{act}$ is related to the chemical activation energy of the weathering reaction and is set to 0.09, and $k_\text{run}$ is a runoff efficiency factor set to 0.045, which helps account for changes in precipitation. $\beta$ is the dependence on pCO$_2$, and is set to 0.5 \citep{menou2015,pierrehumbertbook,kump00}. We follow a similar procedure to \citet{Edson2012} and \citet{Lehir2009} to compute this locally across the planet and then integrate to compute the global weathering rate. When the planet is not in a snowball state, implying global mean temperatures above approximately 260 K, we evaluate \autoref{eq:carbon} at each cell if the cell has land surface and has surface temperatures above 273.15 K, 4 times per simulated day in 6-hour intervals. The weathering rate is averaged over a full year in each cell, to compute a localized annual mean weathering rate. This is then spatially averaged over the globe to compute the annual mean global weathering rate. 

We run PlaSim in the T21 configuration, with 64 latitudes and 32 longitudes. We use 10 vertical grid levels, turn on the mixed-layer ocean, and use a timestep of 45 minutes. We allow pCO$_2$ to change year-after-year according to the annual weathering and outgassing rates. This change is usually extremely small, but it costs nothing if we are already computing the weathering rate, and yields slightly more accuracy, particularly during episodes of very fast weathering when a planet has just emerged from a snowball state. We run PlaSim for a minimum of 11 years, at which point we use NumPy to perform a linear fit to the annual mean surface temperature. We average the slopes of fits for 9-year histories ending in each of the three most recent years to give a sense of the general trend of the past decade and to try to minimize the effect of outliers. If this average slope has an absolute value greater than 0.05 K/year, then we consider PlaSim to not have reached thermal equilibrium, and allow it to continue to run for one more year. This check is considered at the end of each simulated year. This procedure has been determined empirically to achieve approximate thermal equilibrium, and other methods for determining thermal equilibrium should work as well, achieving similar results.

The zero-dimensional pCO$_2$-evolution component of the model is computed by solving 
\begin{equation}
\frac{d}{dt}pCO_2 = V - W(T_\text{s},pCO_2),
\end{equation}
as in \citet{menou2015}. We first compute the annual temperature change relative to the change in pCO$_2$ over the last two years of the simulation, $\partial T/\partial\text{pCO}_2$. Since the surface temperature is the quantity most directly related to sea ice extent, which determines the qualitative model state, we aim to control the temperature evolution, rather than simply controlling the pCO$_2$ evolution. This is because sea ice extent and surface temperature become increasingly sensitive to perturbations in pCO$_2$ near the global freeze/thaw point. We therefore compute the pCO$_2$ step size as
\begin{equation}
\Delta pCO_2 = \text{sign}(V-W)\Delta T\frac{\partial pCO_2}{\partial T},
\end{equation}
where $\Delta T$ is the absolute value of the desired temperature change, $V$ is the volcanic CO$_2$ outgassing rate, and $W$ is the global CO$_2$ weathering rate. The weathering rate is only needed at this point to determine the sign of the change. In order to gain higher sampling resolution of the transitions between snowball and warm states, we decrease the temperature step size $\Delta T$ near the global freeze/thaw point. For temperatures between 240 and 268 K, we aim for a change of 0.5 K per step. For temperatures between 235--240 K and 268--273 K, we aim for 1 K per step. And for temperatures below 235 K or above 273 K, we aim for 2 K per step. We use finer sampling near the transition because a model with too-coarse sampling could artificially introduce climate cycles in models with weathering equilibrium temperatures near the global freeze/thaw point. 

We can combine this pCO$_2$ step size with the pCO$_2$ rate of change at the end of the last model year to estimate the duration of the pCO$_2$ adjustment:
\begin{equation}\label{eq:co2evol}
\Delta t = \frac{V - W}{\Delta pCO_2}.
\end{equation}
Because this estimate uses only the pCO$_2$ rate of change at the end of the previous relaxation step, this method is likely not very accurate. However, for the purposes of identifying climate cycles and the order of magnitude of their timescales, this method is likely sufficient. A more robust and sophisticated technique could perhaps be used in its place to yield more accurate estimates of the period of a given cycle, but is beyond the scope of this paper.

\subsection{Establishing Limits on Climate Stability}
\label{sec:grid}

We further attempt to determine the limits of the temperate habitable zone, beyond which limit cycles should occur. We accomplish this with a parameter sweep, running PlaSim to thermal equilibrium for 31 different insolations, ranging from 1.0 F$_\oplus$ to 0.75 F$_\oplus$, where F$_\oplus$ is the incident solar flux received by the present-day Earth, taken here to be 1367 W m$^{-2}$. These insolations correspond to orbital radii between 1 and 1.15 AU, separated by 0.005 AU. For each insolation, we prescribe a CO$_2$ partial pressure ranging from 0.1 $\mu$bars to 6 bars, separated by 0.1 dex, and run the model to thermal equilibrium for that given pCO$_2$ and insolation. At the transition points between snowball and temperate climates, we reduce the sample spacing to 0.01 dex. We do this for both warm, Earth-like initial conditions with only limited glaciation, and snowball initial conditions. We use Earth-like land distributions and orography. We compute the global weathering rate averaged over two decades for each model at thermal equilibrium. These values give an estimate of the edges of the temperate habitable zone in terms of outgassing rates and insolation. 

\section{Results}

\begin{figure*}
\begin{center}
\subfigure[]{\includegraphics[width=4in]{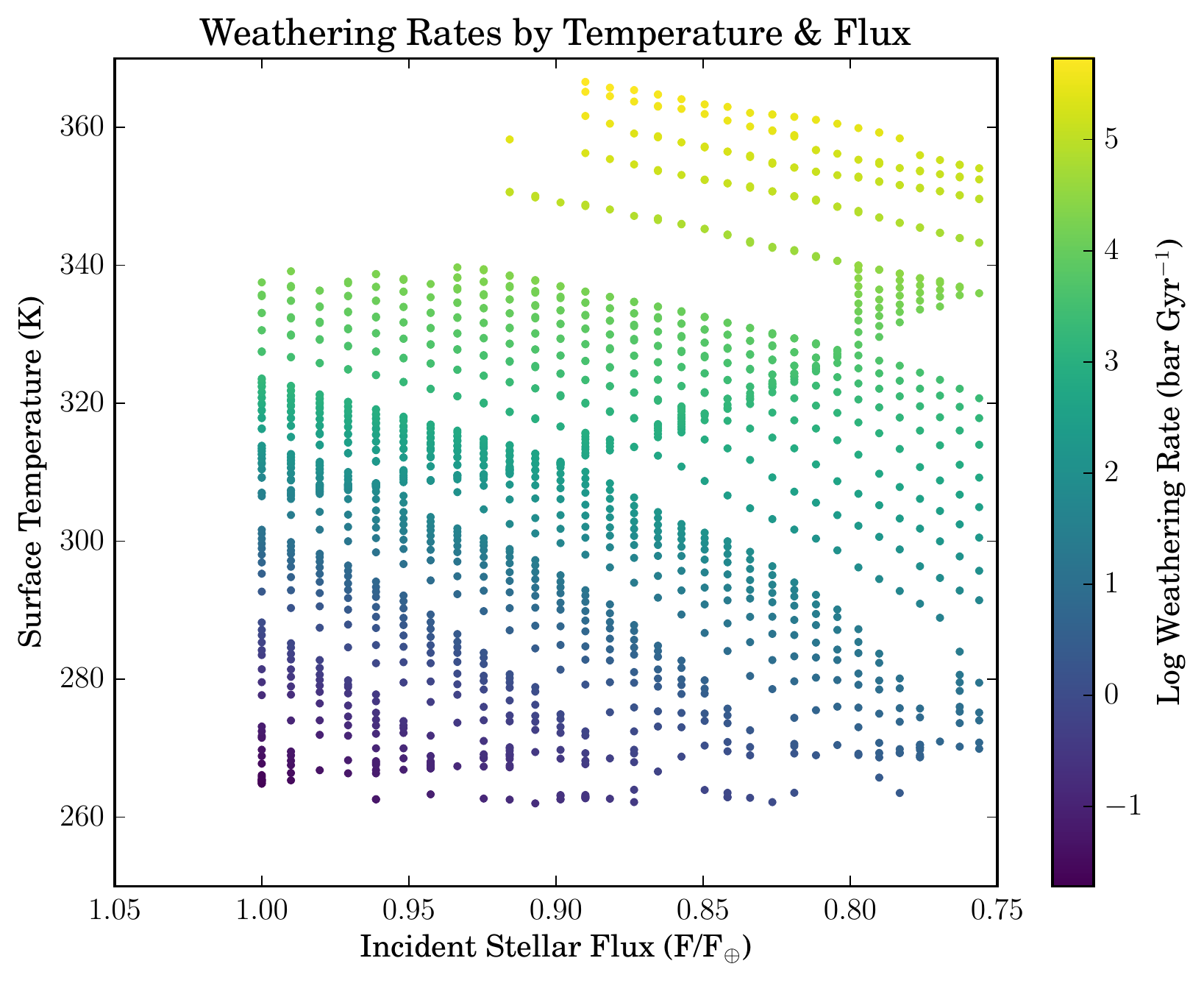}}
\subfigure[]{\includegraphics[width=4in]{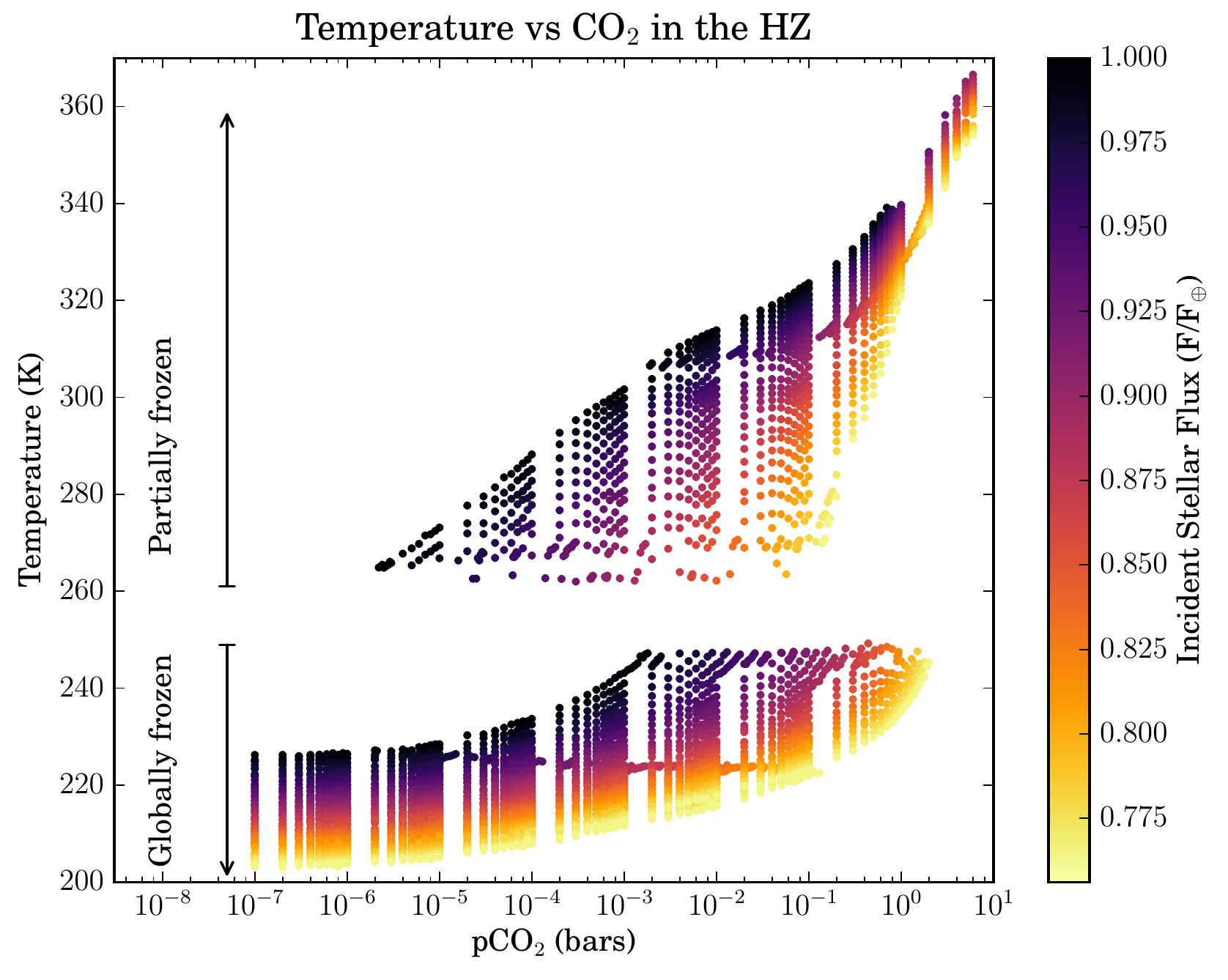}}
\end{center}
\caption{All models run in the parameter sweep, presented in two different ways. Each point represents the annual mean temperature of the last year of a model run to thermal equilibrium for a given CO$_2$ partial pressure and insolation. The figure in (a) shows the weathering rate at thermal equilibrium for each temperate climate, excluding models with no weathering (snowball states), and (b) shows the dependence on pCO$_2$, which illustrates the insolation bistability proposed by \citet{Budyko1969}. No thermal equilibria exist for temperatures between approximately 250--260 K, due to an ice-albedo feedback process that is unstable to perturbations. Models on either side of this gap therefore represent the globally frozen states that comprise the majority of the cycle duration, and the partially frozen or globally warm states which punctuate the cycles. The weathering rates of the coldest models in (a) represent the minimum outgassing rates for temperate stability.}
\label{fig:tempsweep}
\end{figure*}

\begin{figure}
\begin{center}
\includegraphics[width=4in]{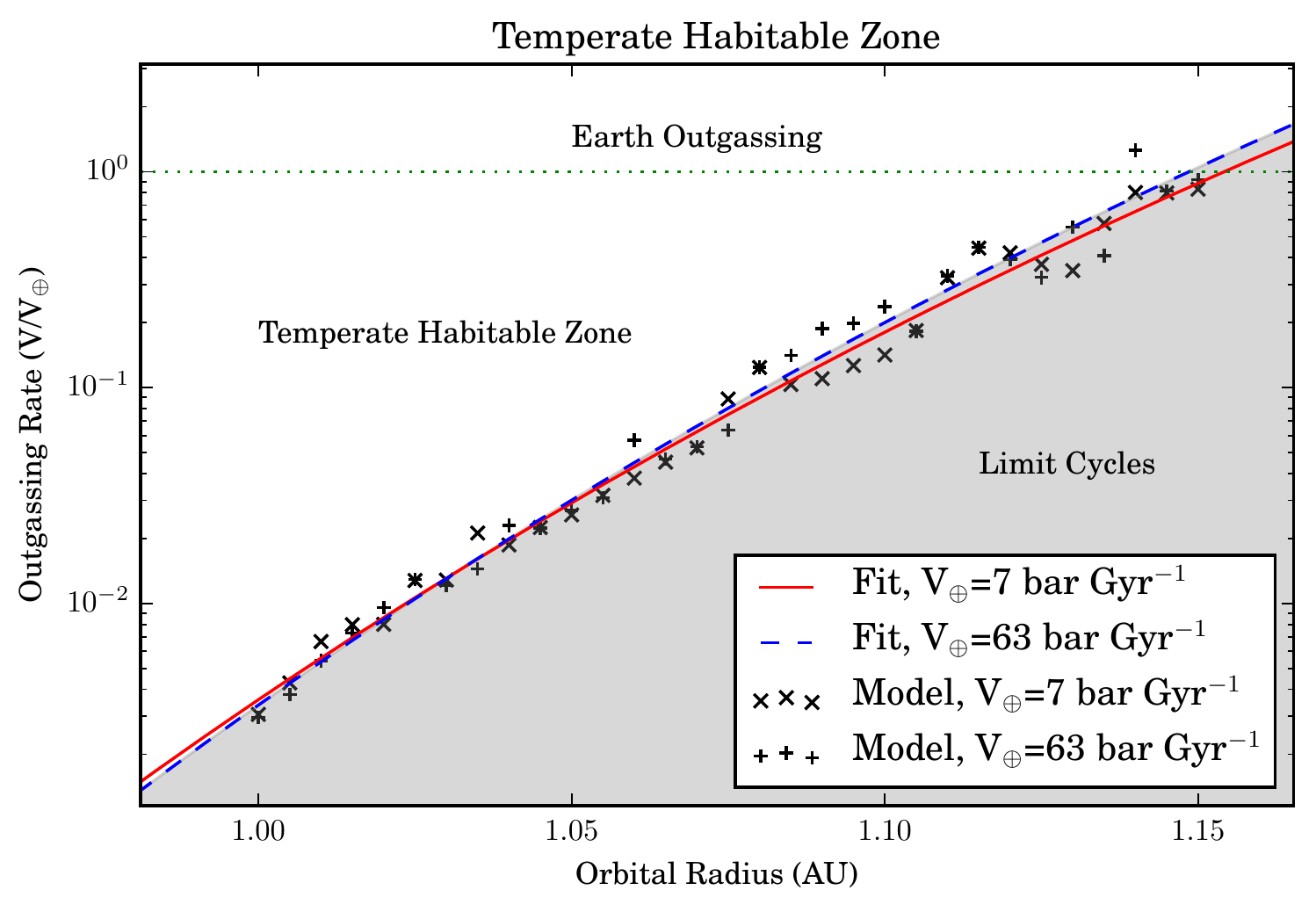}
\end{center}
\caption{Minimum outgassing rates for stable temperate climates as a function of insolation and the decomposition of the habitable zone into temperature and limit cycling regions. \citet{haqqcycles} suggested that 63 bars Gyr$^{-1}$ may be a more appropriate estimate for Earth's outgassing rate than the 7 bars Gyr$^{-1}$ used in \citet{menou2015}, so we compare results assuming each of the two rates. There is no effect on the limit of the temperate habitable zone; the only effect is on limit cycle timescales. Both obey an exponential trend, such that $v=\kappa e^{-\alpha f}$, where the outgassing $v=V/V_\oplus$, and insolation $f=F/F_\oplus$. In this fit, $\kappa\approx3.6\times10^7$, and $\alpha\approx23$. We show the data (black) and fits (red and blue) for both assumptions, normalized to Earth in each case, showing the strength of the agreement. Therefore, the habitable zone can be divided into a temperate habitable zone, consisting of outgassing rates above the fit, and limit cycles beyond that. For modern Earth, that suggests that the temperate habitable zone extends to approximately 1.14--1.15 AU.}
\label{fig:transition}
\end{figure}

The parameter sweep, whose results are shown in \autoref{fig:tempsweep}, show a range of temperatures between roughly 250 and 260 K for which no equilibriated models are found. This is because in that temperature range, sensitive albedo feedbacks with sea ice lead to either rapid glaciation into a snowball state or a rapid thaw into a temperate state. The weathering rates at the lower limit of the partially-frozen models are therefore the minimum outgassing rates required to maintain stable temperate climates. These rates are shown in \autoref{fig:transition}. The minimum outgassing rate is exponentially related to the insolation, such that
\begin{equation}\label{eq:fit}
\frac{V}{V_\oplus} = \kappa e^{-\alpha\left(\frac{F}{F_\oplus}\right)},
\end{equation}
where $\kappa$ and $\alpha$ are constants, and in our fit equal to approximately $3.6\times10^7$ and 23 respectively. This has a similar functional form to the limit of the temperate habitable zone derived by \citet{Abbot2016}. \citet{haqqcycles} argued that \citet{menou2015} used an inappropriate estimate of Earth's outgassing rate, citing a more appropriate value of approximately 63 bars Gyr$^{-1}$. We test the model's dependence on this by running the same models with 63 bars Gyr$^{-1}$ as the Earth outgassing rate. The results are  shown in \autoref{fig:transition} as a second set of datapoints. Since the weathering parameterization, \autoref{eq:carbon}, uses values normalized to Earth, changing the Earth outgassing rate doesn't affect the onset of limit cycling relative to Earth outgassing, instead affecting the timescales of the cycles. The result is that climate cycles occur at actual outgassing rates approximately 10 times higher than suggested when assuming Earth outgasses 7 bars Gyr$^{-1}$, and at timescales approximately 10 times shorter.

\begin{figure*}
\begin{center}
\includegraphics[width=6in]{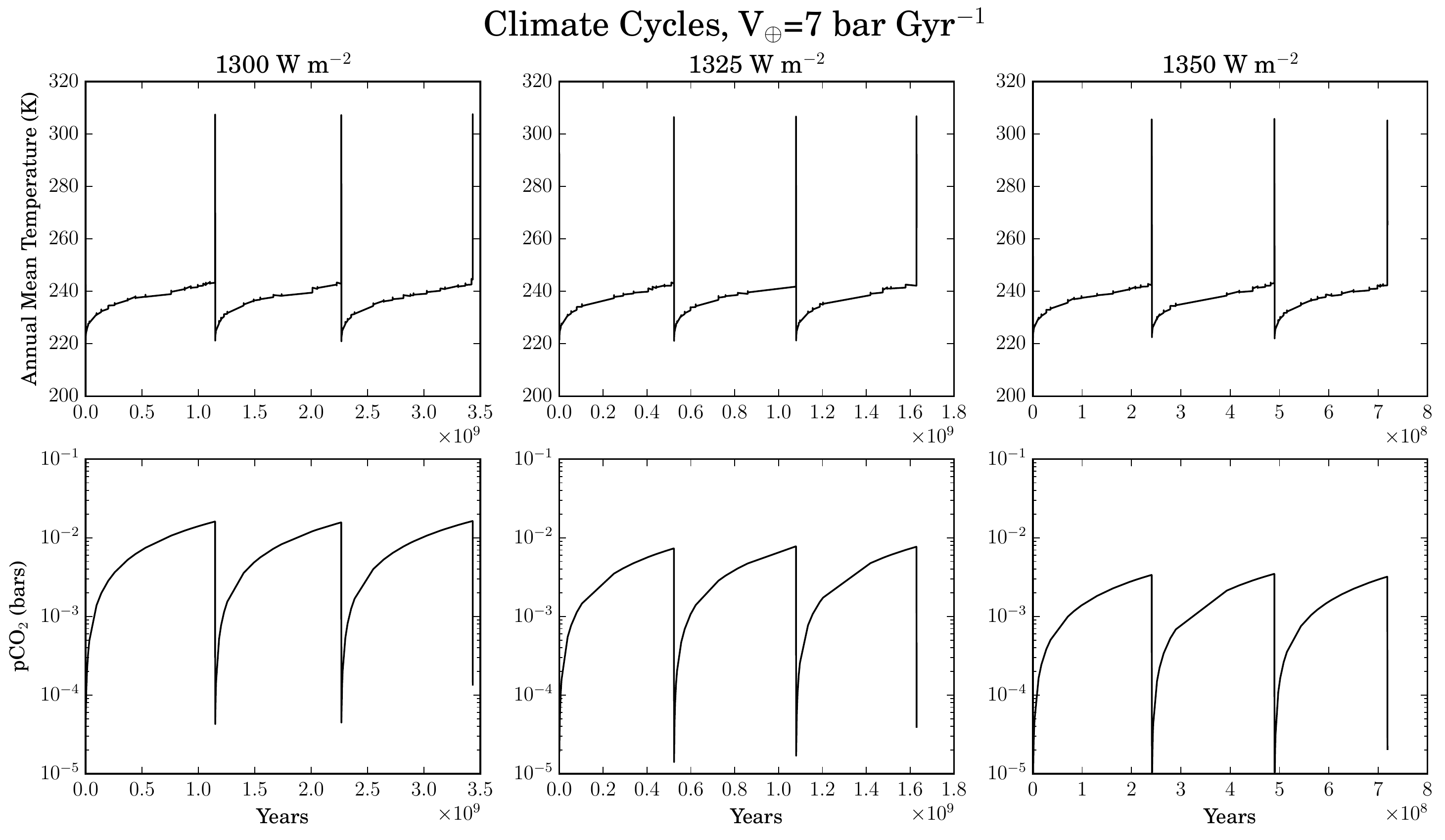}
\end{center}
\caption{The climate cycles observed at 1300, 1325, and 1350 W m$^{-2}$ insolation, with $2\times10^{-3}$ V$_\oplus$. These insolations correspond to approximately 1.006, 1.016, and 1.025 AU in our own solar system. The outgassing rate was chosen to be in the unstable regime at 1350 W m$^{-2}$, as shown in \autoref{fig:transition}, and these three insolations were chosen to probe climate behavior throughout limit cycles without reaching the higher pCO$_2$ levels that would be necessary for deglaciation at lower insolations. The cycles have a period of  hundreds of Myrs to 1 Gyr, and the mean temperature and pCO$_2$ curves have the same form as those identified in lower-dimensional models in \citet{menou2015,haqqcycles}. The cycle periods decrease by roughly a factor of 10 if Earth is assumed to have a higher outgassing rate, as in \citet{haqqcycles} and \citet{Abbot2016}.}
\label{fig:cycles}
\end{figure*}

\begin{figure*}
\begin{center}
\includegraphics[width=6in]{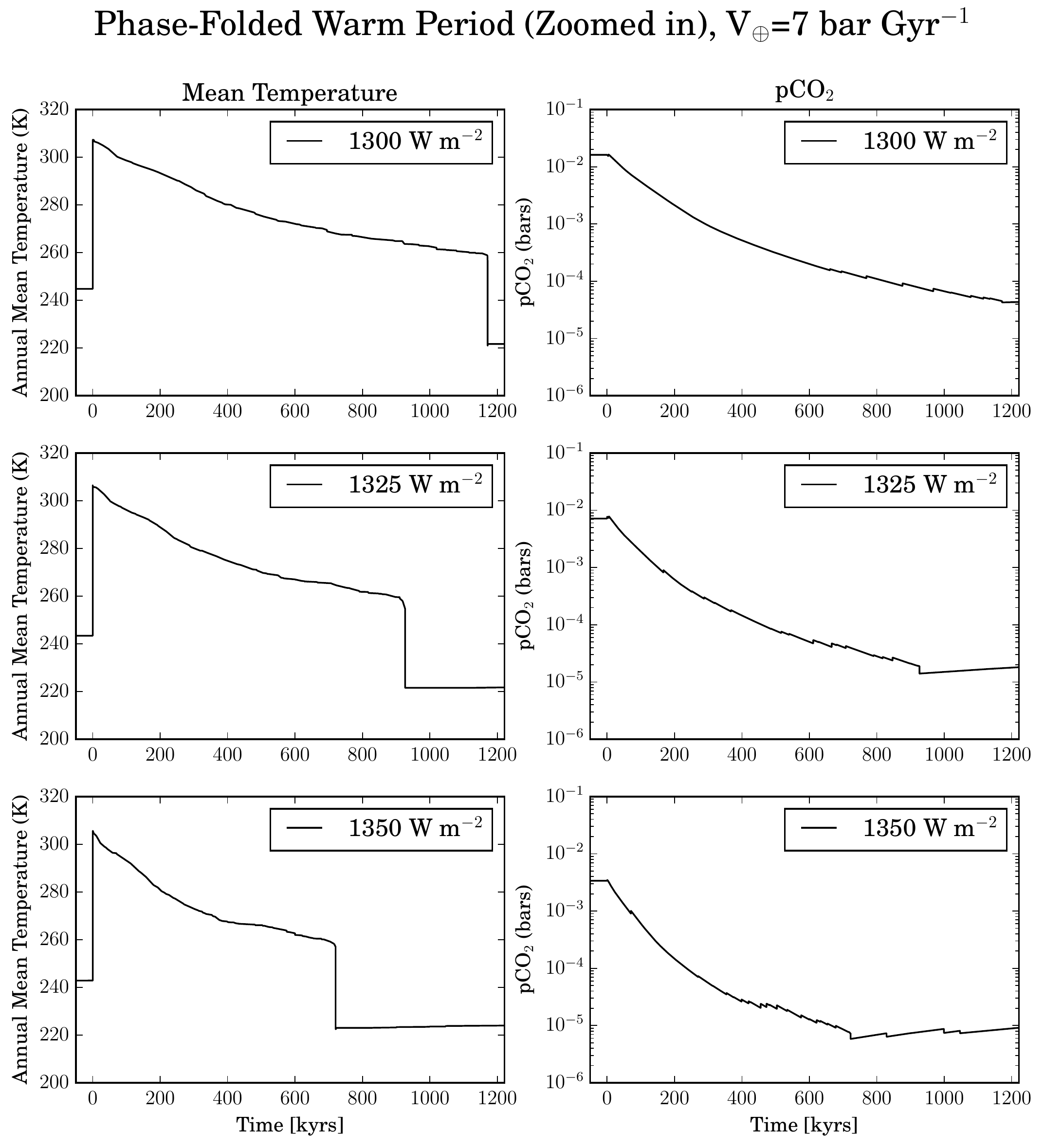}
\end{center}
\caption{The warm periods of the cycles shown in \autoref{fig:cycles}, phase-folded to increase the resolution of the warm branch and therefore increase the accuracy of the timescale estimate. These periods last on the order of hundreds of thousands to a million years. Note that much of the reduction in pCO$_2$ happens rapidly, with the climate cooling steadily until a tipping point is reached. The cooldown during the temperate phase involves nonlinear and highly local sea ice effects, but because it occurs on near-Myr timescales, still requires only intermittent GCM steps. Compared to the relatively uneventful snowball outgassing phase, capturing enough of the nonlinearity to describe the cooldown requires striking a balance between time-sampling resolution and computational expense. This is why we simulate several cycles, and then phase-fold them to increase the time-sampling resolution and reduce the likelihood of an outlier timestep affecting the timescale estimate.}
\label{fig:dutycycle}
\end{figure*}

Our hybrid 0D-3D models were run at $2\times10^{-3}$ V$_\oplus$, or 14 mbar Gyr$^{-1}$, at 1350 W m$^{-2}$, 1325 W m$^{-2}$, and 1300 W m$^{-2}$ insolation. We assume modern Earth obliquity and seasons, orbital eccentricity, land fraction, and continental distribution and orography. As expected, we find that these models have unstable climates characterized by limit cycling, as shown in \autoref{fig:cycles}. The cycles have periods ranging from hundreds of millions to approximately 1 billion years. We phase-fold the warm periods to improve their resolution and increase the accuracy of our estimate of their average duration, finding that they last for approximately 1.2 Myr, 930 kyr, and 720 kyr respectively, as shown in \autoref{fig:dutycycle}. The warm periods are therefore of order 0.1\% of the total cycle duration at these insolations and outgassing rates. It is also of note that the mean temperature and pCO$_2$ curves throughout the cycle have the same form as those identified in \citet{menou2015}. Inconsistencies between different cycle periods are likely the result of imperfections in the pCO$_2$-stepping algorithm and estimation of the time intervals between data points. With an improved model and algorithm for spanning these geological timescales, we expect greater consistency between cycles. Using the outgassing estimate in \citet{haqqcycles}, we find that cycle periods decrease by roughly a factor of 10, consistent with a nearly factor of 10 increase in the speed at which atmospheric pCO$_2$ is replenished. These are consistent with the cycle periods found by \citet{Abbot2016}.

To probe the sensitivity to continental distribution, we run a set of models at Earth insolation and 21 mbar Gyr$^{-1}$ outgassing, or 0.3\% V$_\oplus$. Each has the same land fraction as modern Earth, but the land is distributed either in a rectangular supercontinent centered on the equator, a circular supercontinent centered on a pole, two circular continents centered on the poles, and three models with most land at the poles but with a small continent of varying size at the equator. Each model assumes flat orography for simplicity. Of these models, only the equatorial supercontinent demonstrates an unstable climate at this insolation and outgassing, suggesting a strong sensitivity to continental distribution, consistent with the findings in \citet{Donnadieu2006} that weathering rates are strongly-affected by changes in land distribution, with tropical continents leading to higher weathering rates.

\begin{figure}
\centerline{
\subfigure[6.5 $\mu$bars CO$_2$]{\includegraphics[width=0.33\linewidth]{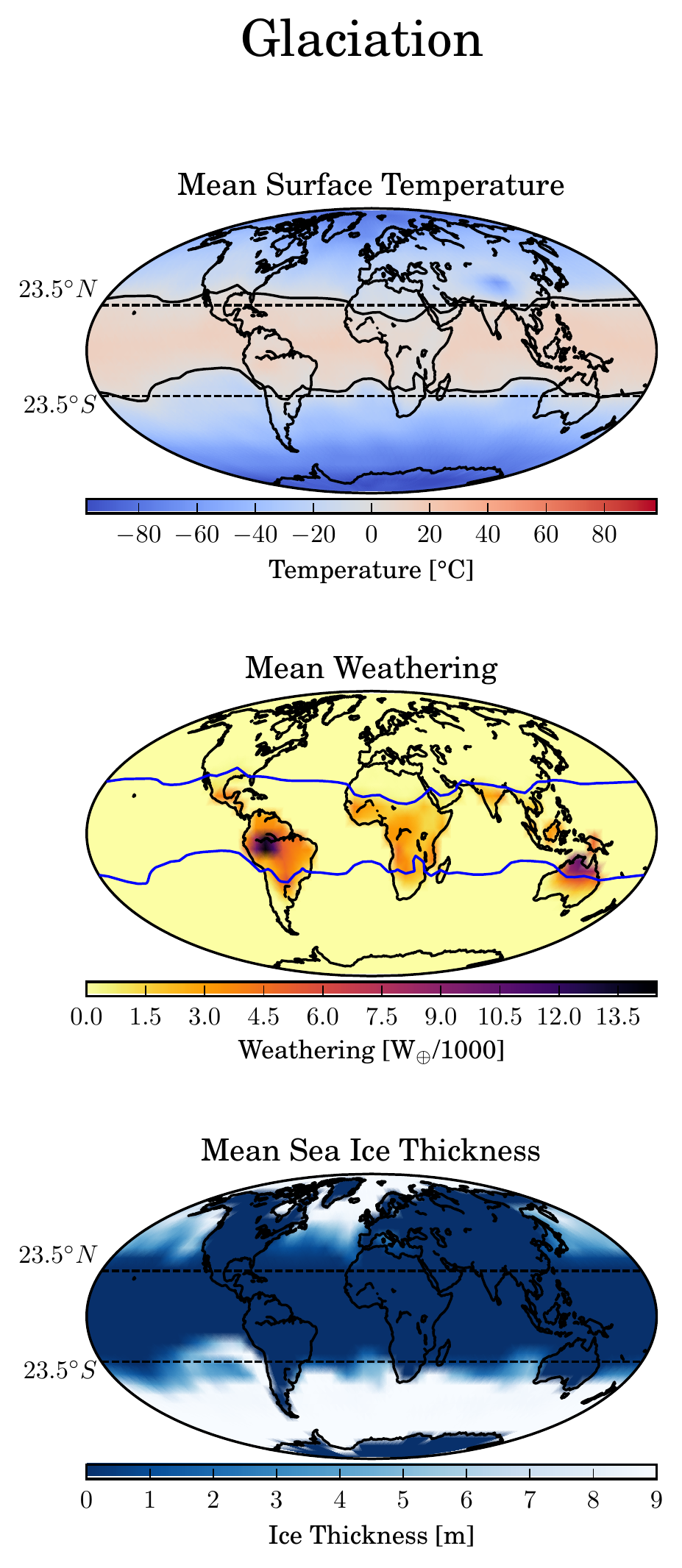}}
\subfigure[6.6 $\mu$bars CO$_2$]{\includegraphics[width=0.33\linewidth]{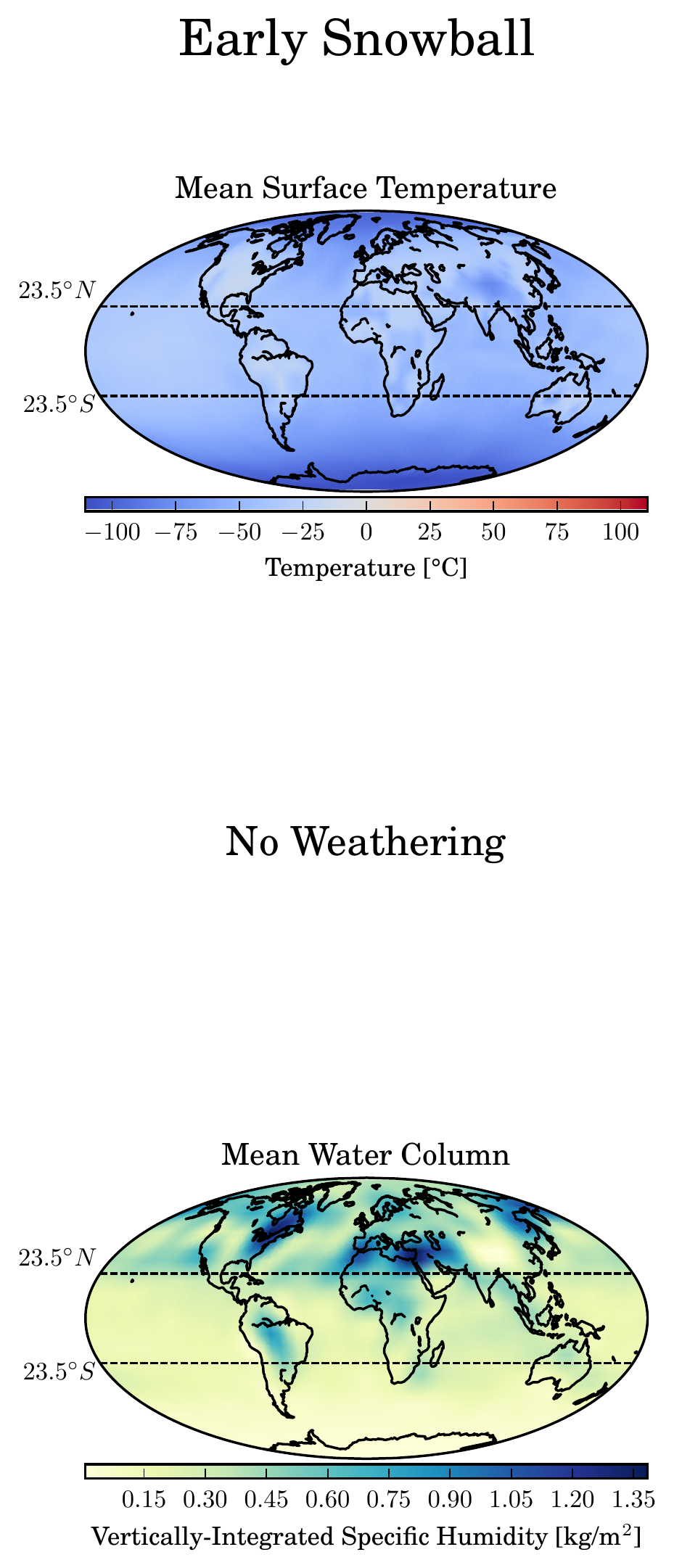}}
\subfigure[3.4 mbars CO$_2$]{\includegraphics[width=0.33\linewidth]{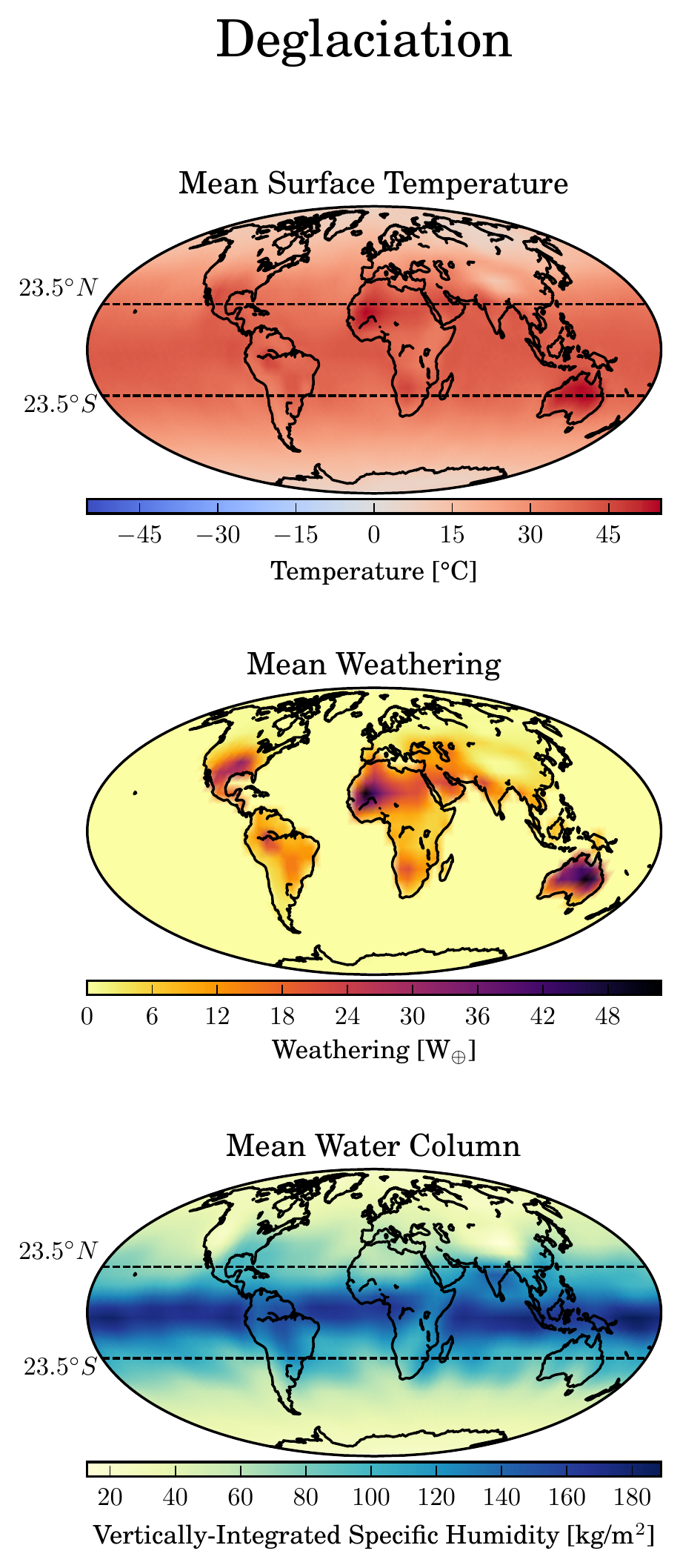}}
}
\caption{Three critical points in the snowball hysteresis at 1350 W/m$^2$ insolation are shown: the glaciation point when the climate is about to transition to a snowball state, the initial snowball state following that transition, and the deglaciated climate following the transition back out of snowball. These climate states are found in our simulations at pCO$_2$ levels of 6.6 $\mu$bars (6.5 ppmv), 6.5 $\mu$bars (6.4 ppmv), and 3.4 mbars (0.3\%), respectively. Columns (a), (b), and (c) depict annual averages for each grid point. The glaciation point is characterized by sea ice extending roughly to the Tropics of Cancer and Capricon at $\pm23.5$\degree  latitude, and weak weathering concentrated on tropical landmasses. The 0 \degree{C} surface isotherm is shown as a solid black line on the temperature plot in column (a) and as a solid blue line on the weathering plot in column (a). The initial snowball state is characterized by globally cold temperatures, reaching lows of -156 \degree{C}, tropical snowfall, and very low specific humidity with water vapor primarily at mid- and high latitudes, as shown in column (b). The deglaciation exit point, when all sea ice is melted, is conversely characterized as being extremely warm, with average temperatures above-freezing everywhere and highs peaking at 66 \degree{C}, very strong weathering, and very high specific humidity at tropical latitudes, as shown in column (c).}
\label{fig:maps1}
\end{figure}

\begin{figure}
\centerline{
\includegraphics[width=1.0\linewidth]{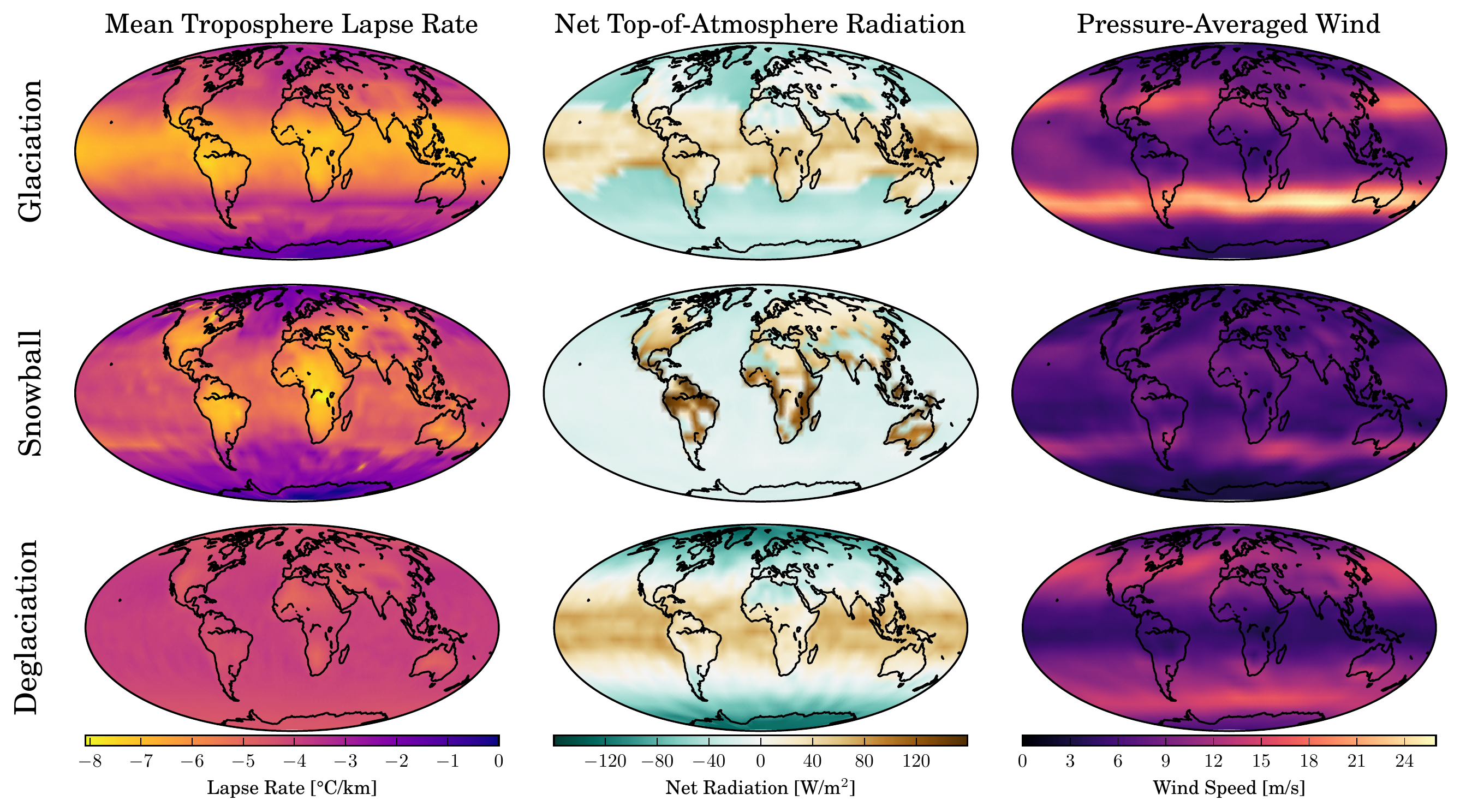}
}
\caption{Tropospheric lapse rate, net top-of-atmosphere radiation (radiation down minus radiation up), and pressure-averaged wind speed for each of three critical points in the snowball hysteresis at 1350 W/m$^2$: the glaciation point where the temperate climate is on the verge of tipping into a snowball state, the snowball state that climate arrives in, and the deglaciated state immediately after the exit from the snowball state. The pressure-averaged wind speed is the average of the wind speed at each pressure level, weighted by the air pressure at that level. A stronger average wind speed therefore generally means stronger winds at low altitudes (high pressures). The glaciation critical point shows strong latitudinal variation in the lapse rate, a boundary between top-of-atmosphere heating and cooling defined by the sea-ice boundary, and strong low-altitude wind streams at mid-latitudes. The initial snowball state, meanwhile, has moderate lapse rate gradients between land and sea ice, net warming over land, and generally weak atmospheric circulation. The deglaciation critical point, however, shows near-uniform lapse rates, tropical top-of-atmosphere heating and polar cooling primarily defined by the insolation shape, and high-altitude wind streams at higher latitudes that are generally more diffuse.}
\label{fig:maps2}
\end{figure}

\section{Simulation Caveats}\label{sec:caveats}
\subsection{Model Uncertainties}\label{subsec:model}

\begin{figure}
\begin{center}
\includegraphics[width=6in]{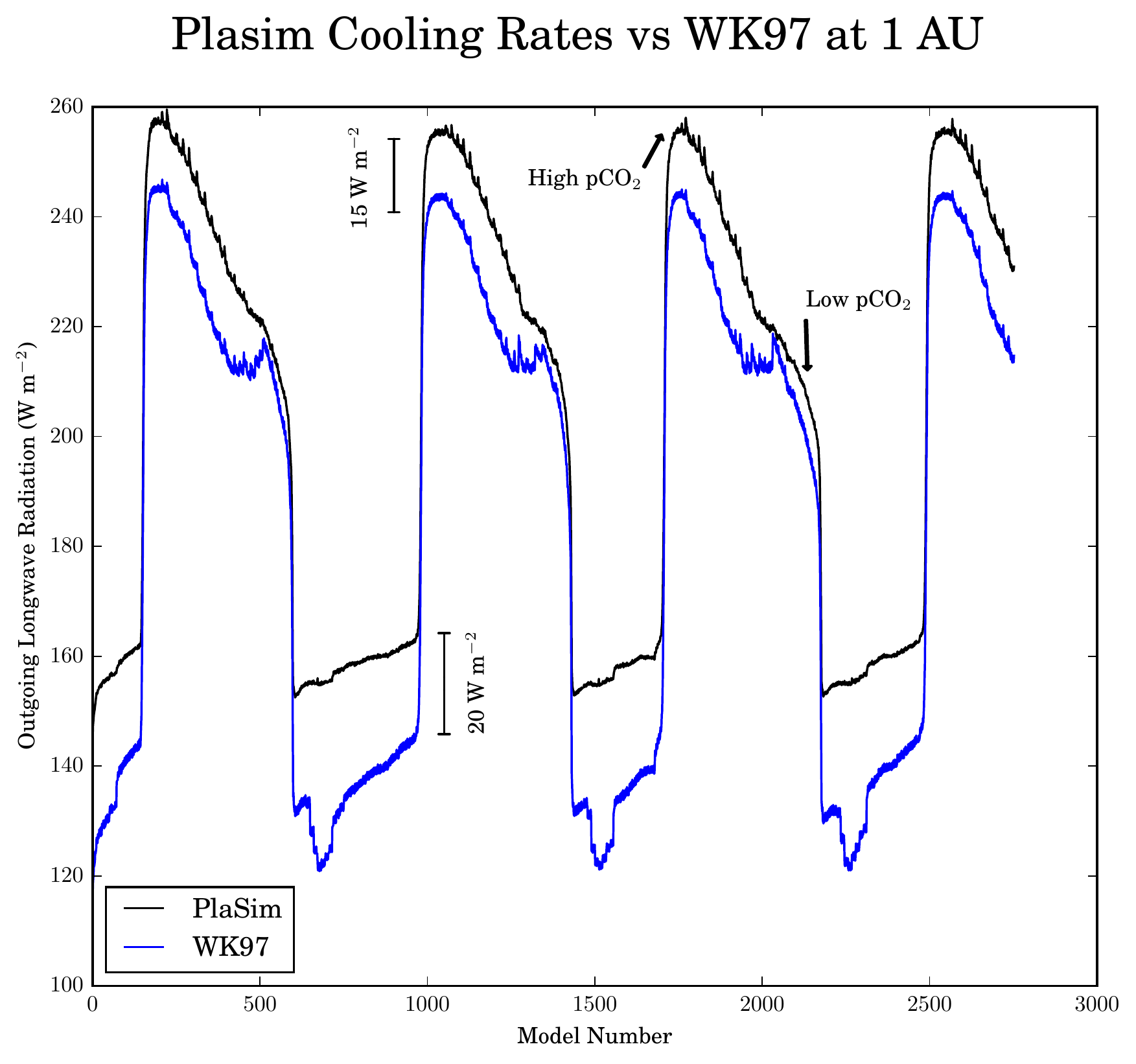}
\end{center}
\caption{Comparison of outgoing longwave radiation (OLR) throughout cycles at 1 AU and mbar Gyr$^{-1}$ outgassing as computed by PlaSim, and as computed by the polynomial fit proposed by \citet{wk97}. Each model number is one GCM year. We compute the WK97 polynomial by computing the global average surface temperature over the 3D GCM outputs, and compute a single polynomial with that and pCO$_2$ as input. The two models predict global cooling rates that can differ significantly in the high-pCO$_2$ regime, by up to tens of percent. This suggests substantial uncertainty in the quantitative abundance and duration of limit cycles in the habitable zone.}
\label{fig:wk97}
\end{figure}

While PlaSim represents a model of much greater complexity than the simplified model used in \citet{menou2015}, it does have limitations. The radiation model used in PlaSim is highly simplified, largely ignoring aerosols and finer spectral variations in absorptivities, relying on broadband emmisivities, polynomial prescriptions, and constant albedos, and insufficiently treating the stratosphere, resulting in biases \citep{plasimdocs}. For example, different models can predict significantly different cooling rates for a given $T$-pCO$_2$ combination, resulting in different abundances and durations of snowball states. \citet[hereafter WK97]{wk97} ran a suite of 1D models for various pCO$_2$ levels, and then fit polynomials to the radiative results of the models. They claim an rms error in outgoing longwave radiation (OLR) of less than 5 W/m$^2$ in the 190--370 K temperature range and from 10$^{-5}$--10 bars pressure range. Almost all of our models fall within this pCO$_2$ range. We compute a global annual mean of outgoing longwave (thermal) radiation by recording the outgoing longwave radiation (OLR) at each grid point 4 times per simulated day. We then compare this with the OLR predicted by the WK97 polynomial fit assuming the same global annual surface temperature and pCO$_2$, computing a single polynomial with the global values as inputs. The comparison is shown in \autoref{fig:wk97}. PlaSim consistently predicts more cooling, by 10\% or less during warm phases, but by up to 70\% in the frozen phase when pCO$_2$ levels are higher. Even in a best-case scenario of Earth insolation, using 21 mbar Gyr$^{-1}$ outgassing to trigger limit cycles, PlaSim predicts more cooling. This shortcoming of the model almost certainly affects the quantiative accuracy of the results. We however consider this loss of accuracy to be one of the tradeoffs to be made in exchange for improved model speed that allows us to simulate long timescales and large swaths of parameter space. There is a balance to be struck, and so the reader should be wary of quantitative estimates pertaining to the high-pCO$_2$ regimes. Thus, without repeating this study with different GCMs, our results generally cannot be considered quantitative predictions for where climate cycles might occur beyond identifying qualitative trends---we expect limit cycles to dominate the outer habitable zone, with a boundary roughly described by an exponential trend, as in \autoref{eq:fit} and as predicted by \citet{Abbot2016}.

PlaSim also neglects glacial dynamics---instead, snow simply accumulates on land. This means that glacial flow is ignored, and any differences in albedo or melting properties between large accumulations of snow and glaciers are ignored. And while PlaSim includes the effect on albedo of snow-covered sea ice as opposed to bare ice, the effect of dust accumulation in snow is not. This could produce qualitatively different results in how the planet melts and freezes. We also ignore any interaction between glaciation and volcanism. There likely is some important interaction, as some studies suggest that rapid changes in lithostatic pressure due to rapid melting could trigger volcanic activity \citep{icevolcano1}, and others suggest that increased lithostatic pressure on continents due to glacial loading could also trigger volcanic activity \citep{icevolcano2}. The amount of volcanic activity will also have an effect on snow and ice albedo by changing the amount of soot and other contaminants in the snow and ice \citep{volcanicalbedo}. 

Our models also do not take into account changes in sea level due to melting or freezing. As weathering rate depends on land fraction, this may have a significant impact. We also do not fully-treat large-scale ocean circulation, which tends to discourage the advance of sea ice, as currents transport heat away from the equator \citep{dynamicocean}. We also ignore variations in fundamental orbital properties such as stellar luminosity and secular variations in the orbit itself, both of which can be expected to vary on relevant timescales \citep{christa,varysun}.

As a result of these approximations and inaccuracies, it is likely that there is significant uncertainty and inaccuracy in the conditions under which PlaSim predicts entry into a snowball state and deglaciation out of it. Since our aim is not to precisely quantify the conditions under which snowballs and limit cycles occur, but rather to verify that they persist as a phenomenon in higher-dimensional models and to roughly outline the parameter space in which they can be expected, we believe this is an acceptable tradeoff for PlaSim's computational speed. However, it is worthwhile to consider how changes in snowball entry and exit conditions would affect our quantitative results. If glaciation into snowball conditions happens at colder temperatures (lower pCO$_2$), then temperate climates should be stable at lower outgassing rates than those predicted by our results. This would generally diminish the amount of parameter space occupied by limit cycles. However, since we are concerned with a logarithmic outgassing parameter space, we do not expect a reduction in glaciation temperature of a few degrees to imply the near-absence of limit cycles from the parameter space. Conversely, if the entry into snowball happens at warmer temperatures, temperate climates would cease being stable at higher outgassing rates, increasing the prevalence of limit cycles. Changes in the deglaciation temperature should have less of an effect on limit cycle prevalence. If deglaciation happens at warmer temperatures (higher pCO$_2$), then the post-deglaciation climate would be significantly warmer with higher weathering rates. Similarly, colder deglaciation would imply a cooler post-deglaciation climate. However, since the dominant factor in whether limit cycles occur is whether a cooling climate is able to reach a weathering equilibrium before glaciation into snowball, changing the deglaciation point leaves the parameter space unchanged. However, changes to either transition temperature could significantly affect the duration of each part of the cycle. Colder glaciation would increase the length of each phase due to the longer weathering and outgassing times needed for the larger shifts in pCO$_2$, as would warmer deglaciation. Conversely, warmer glaciation and cooler deglaciation would reduce the weathering and outgassing times needed.

\subsection{Weathering Uncertainties}

The carbon-silicate cycle implementation is itself a simplification, limited by parameterizations of Earth weathering, and ignoring mantle CO$_2$ processes. One important simplication is in precipitation and runoff. \autoref{eq:carbon} includes a term that depends on runoff efficiency and temperature. This is intended to account for changes in precipitation and runoff and the resulting effect on weathering \citep{abbot2012}. However, this may miss complexities arising from glacial melt and changes in the hydrologic cycle due to large changes in ice fraction. As ice sheets advance to the equator, the amount of precipitable water present in the atmosphere is expected to dramatically decline, leading to a reduction in weathering due to precipitation and runoff. Our implementation also involves computing what was originally a global, 0D parameterization at each grid point and then taking the average, rather than taking into account such variables as the precipitation specific to that gridpoint. It is important to note that we are effectively taking the average of a polynomial, rather than computing a polynomial with an average input. A more sophisticated weathering model might yield quantitatively different results when used with our long-time evolution technique. Nor do we consider the impact of CO$_2$ condensation on weathering feedbacks. \citet{Turbet2017} showed that snowball states at low insolation may result in large amounts of CO$_2$ deposition as dry ice at the planet's pole. We ignore the effect that could have on limit cycles.

\begin{figure}
\begin{center}
\includegraphics[width=4in]{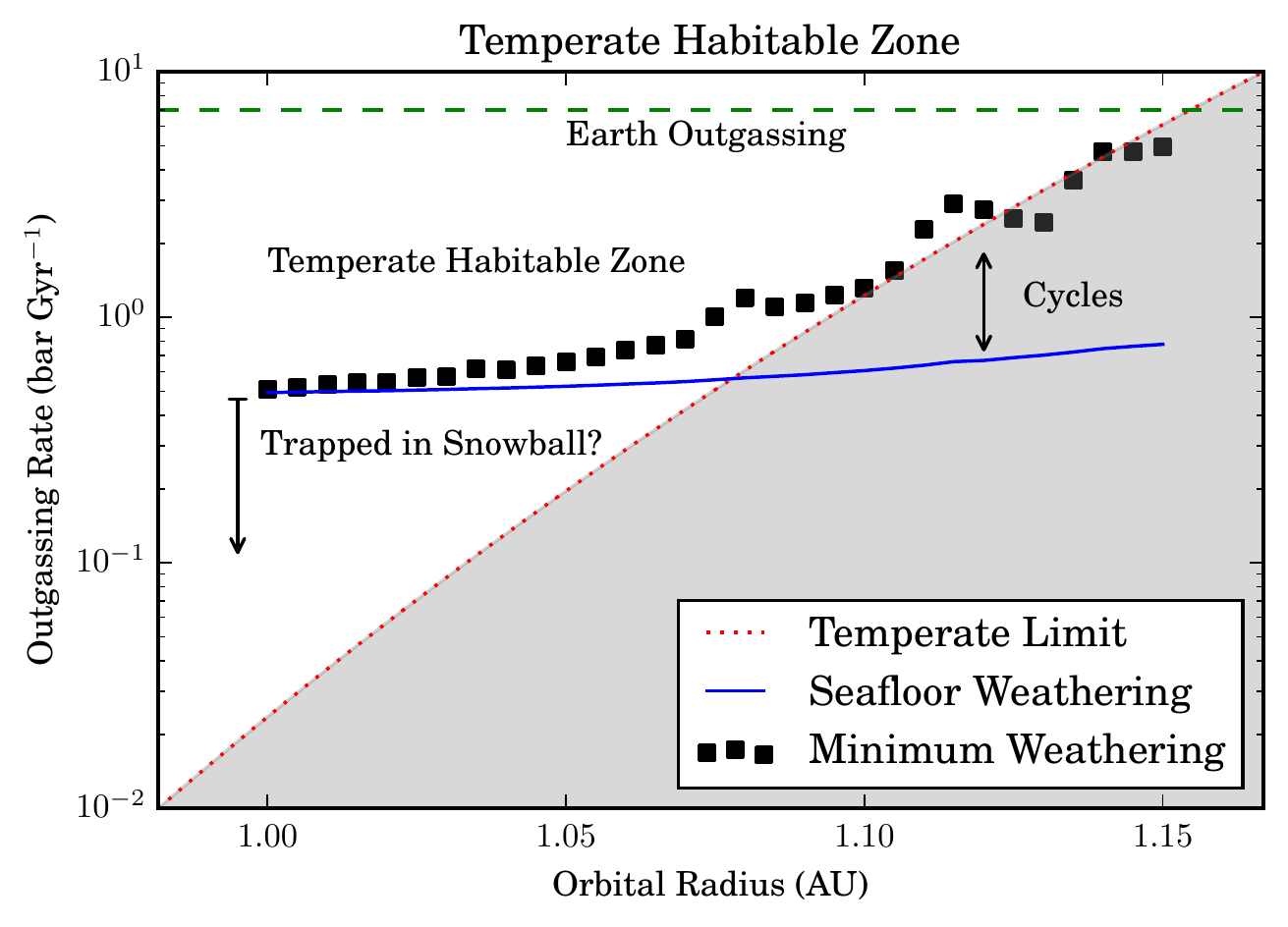}
\end{center}
\caption{The same as \autoref{fig:transition}, with $V_\oplus=7$ bar/Gyr, except here we compute the seafloor weathering rate following \citet{abbot2012} for the pCO$_2$ levels found at the cold end of the temperate branch of the snowball bistability, and add those weathering rates to the minimum sustainable weathering rates computed by the GCM. The result is plotted as squares. If seafloor weathering is included, this sets a floor to the temperate outgassing rates for high insolations, possibly permitting planets to become trapped in snowball states below those limits. The red dotted line represents the limit of the temperate habitable zone assuming no seafloor weathering, and the blue solid line represents the assumed seafloor weathering at that limit.}\label{fig:seafloor}
\end{figure}

We ignore seafloor weathering in our parameterization, as our interest is in the role that continental weathering can play in long-term climate stability, and the role that seafloor weathering plays as a climatic feedback is highly uncertain, as it is not well-known how much carbon is sequestered through that process, nor whether the seafloor weathering rate depends on heat flux from the mantle or the atmospheric temperature \citep{Caldeira1995,abbot2012}. We can however estimate what its impact on our results might be by assuming a parameterization of the same form as that used in \citet{abbot2012}, adopting their notation of $\phi$ as the CO$_2$ partial pressure relative to modern Earth, $pCO_2/pCO_{2,\oplus}$:
\begin{equation}\label{eq:seafloor}
\frac{W_{sf}}{W_\oplus} = \frac{\phi^n+\nu}{1+\nu}
\end{equation}
where $W_{sf}$ is the seafloor weathering, $\nu$ is related to the seafloor spreading rate and is taken to be 12.6, and $n=\frac{1}{3}$ \citep{abbot2012}. The total contribution of seafloor and continental weathering rates is estimated to be a linear combination,
\begin{equation}\label{eq:abbotw}
W = \beta_0W_\text{c} + (1-\beta_0)W_{sf},
\end{equation}
where $\beta_0$ is the relative modern contribution of continental weathering, and $W_c$ is the continental weathering rate. This is generally not going to be constant across a wide range of climates, but in the absence of good constraints, we assume it is, and set it to be 0.75 \citep{abbot2012}. The results are shown in \autoref{fig:seafloor}, and suggest that seafloor weathering could provide a weathering floor at high insolations, such that at outgassing rates less than the seafloor weathering rate planets could become trapped in snowball conditions. This regime of parameter space however already assumes very low outgassing rates, and this particular seafloor weathering parameterization does not significantly affect the shape of the stability parameter space at lower insolations and higher outgassing rates, where it is more likely to be relevant for observed Earth-like exoplanets.

The result of these caveats is that fully modeling a planet with any predictive accuracy is difficult, requires a great deal of physics and computation, and suggests both breadth and diversity in the habitable zone exoplanet parameter space. However, by limiting the investigation to a particular slice of that parameter space, we may investigate the effects of a particular mechanism on climate. Our results suggest that the carbon-silicate feedback mechanism permits climate cycles at low insolations and moderate-to-low outgassing rates, in general agreement with the results of \citet{Tajika2007}, \citet{menou2015}, and \citet{haqqcycles}.

\section{Conclusion}

We present a technique for simulating the evolution of climate systems on geological timescales by using a 3D GCM as a relaxation step in a 0D carbon evolution model. We use this to present the first 3D GCM simulations of the climate cycles predicted by 1D models in \citet{menou2015} and \citet{haqqcycles}. This shows that climate cycles are not limited to simple EBMs, and strengthens the argument for their relevance to real planetary systems. We find that Earth analogs undergo climate cycles at moderate outgassing levels and low insolations comparable to early Earth. Further work includes refining the model assumptions, accounting for other physical processes such as glacial dynamics, seafloor weathering \citep{haqqcycles}, and variable outgassing rates in response to glacial loading. We note that our technique is not specific to PlaSim, and any GCM could be used in its place using the same technique. We also note that we have not mapped the entirety of the parameter space proposed by our models to exhibit climate cycles; we have only probed a few points in that space to find climate cycles. Future work should map the cycle frequencies and characteristics in this region and compare with the results presented in \citet{menou2015}, \citet{haqqcycles}, and \citet{Abbot2016}. It may be additionally useful to explore more sophisticated weathering models which capture more of the relevant physical processes, rather than merely using a 3D temperature field to compute a global average weathering rate.

\section{Acknowledgments}
\acknowledgments
The authors would like to thank Professor Ray Pierrehumbert for his very helpful comments, insights, and suggestions, particularly pertaining to glaciation and deglaciation conditions and prior work on coupling GCMs to processes that happen on longer timescales. The authors would also like to thank Michael Way at the NASA Goddard Institute for Space Studies for his helpful comments on the ROCKE-3D GCM and the dependence of the habitable zone on rotation period. AP is supported by a Centre for Planetary Sciences Graduate Fellowship at the University of Toronto, Scarborough, and by the Department of Astronomy \& Astrophysics at the University of Toronto, St. George. KM is supported by the Natural Sciences and Engineering Research Council of Canada. Computing time was provided by the Canadian Institute for Theoretical Astrophysics at the University of Toronto, St. George. Many thanks are given to my peers for their support. 

\software{PlaSim \citep{Fraedrich2005}, matplotlib \citep{Hunter:2007}} 

\appendix

\section{Computational Cost}\label{appendix1}

Earth-system general circulation models are often very computationally expensive. PlaSim is one of the faster GCMs available. On a compute node with two Intel\textregistered$\,$ Xeon\textregistered$\,$ E5310 quad-core CPUs, clocked to 1.60 GHz, running PlaSim in parallel with 8 threads takes 162 seconds of wall-time on average per simulated year, where wall-time is the real time that a user must wait for a job to finish. On a compute node with 2 octa-core Intel\textregistered$\,$  Xeon\textregistered$\,$ E5-2640 CPUs, clocked at 2.00 GHz, running PlaSim in parallel with 16 threads takes 62 seconds of wall-time on average per simulated year. Overall, PlaSim takes approximately 1300 CPU-seconds per simulated year on the E5310 cores and 1000 CPU-seconds per simulated year on the E5-2640 cores. As an example of a more sophisticated GCM, the ROCKE-3D Earth GCM developed at the NASA Goddard Institute for Space Studies can compute a simulated year in as little as 11 CPU-hours (38,000 CPU-seconds) on their Xeon\textregistered$\,$ E5-2697 nodes using 44 cores \citep{NASAGCM}. Climate cycles, as proposed by \citet{menou2015}, occur on the order of hundreds of millions of years to billions of years. Taking as an optimal computing case, if we assume a climate cycle could be found which lasts only 100 Myr, then on the computing hardware available to us, computing the entire cycle directly year-by-year with PlaSim on 16 cores would take $6.2\times10^9$ seconds of wall-time, or approximately 196 years. Since submitting a computing job with a run-time of two centuries is not practical or desirable, directly simulating the carbon-silicate cycle with a GCM is impossible without considerable computing resources. Our technique, however, allows us to avoid directly computing each year of climate evolution, at the cost of computational accuracy. Instead, we compute only a few decades intermittently throughout the long-term evolution of the planet. Computing two periods of the cycles observed in this paper involved directly simulating approximately 1800 years, which is 6 orders of magnitude less than directly computing the entire cycle, and allows full computation in only 31 hours of wall-time on our hardware. 


\begin{thebibliography}{}
\expandafter\ifx\csname natexlab\endcsname\relax\def\natexlab#1{#1}\fi

\bibitem[{Abbot(2016)}]{Abbot2016}
Abbot, D.~S. 2016, The Astrophysical Journal, 827, 117

\bibitem[{{Abbot} {et~al.}(2012){Abbot}, {Cowan}, \& {Ciesla}}]{abbot2012}
{Abbot}, D.~S., {Cowan}, N.~B., \& {Ciesla}, F.~J. 2012, \apj, 756, 178

\bibitem[{Abbot {et~al.}(2010)Abbot, Eisenman, \& Pierrehumbert}]{Abbot2010}
Abbot, D.~S., Eisenman, I., \& Pierrehumbert, R.~T. 2010, Journal of Climate,
  23, 6100

\bibitem[{{Batalha} {et~al.}(2016){Batalha}, {Kopparapu}, {Haqq-Misra}, \&
  {Kasting}}]{earlymars}
{Batalha}, N.~E., {Kopparapu}, R.~K., {Haqq-Misra}, J., \& {Kasting}, J.~F.
  2016, Earth and Planetary Science Letters, 455, 7
  
\bibitem[{Berner(1994)}]{Berner1994}
Berner, R.~A. 1994, American Journal of Science, 294, 56

\bibitem[{Berner(2001)}]{Berner2001}
---. 2001, American Journal of Science, 301, 182

\bibitem[{Berner(2004)}]{Berner2004}
---. 2004, The Phanerozoic Carbon Cycle: CO$_2$ and O$_2$ (New York: Oxford
  University Press)

\bibitem[{Boschi {et~al.}(2013)Boschi, Lucarini, \& Pascale}]{Boschi2013}
Boschi, R., Lucarini, V., \& Pascale, S. 2013, Icarus, 226, 1724

\bibitem[{Budyko(1969)}]{Budyko1969}
Budyko, M.~I. 1969, Tellus, 21, 611

\bibitem[{Caldeira(1995)}]{Caldeira1995}
Caldeira, K. 1995, American Journal of Science, 295, 1077

\bibitem[{Donnadieu {et~al.}(2006)Donnadieu, Godd{\'{e}}ris, Pierrehumbert,
  Dromart, Jacob, \& Fluteau}]{Donnadieu2006}
Donnadieu, Y., Godd{\'{e}}ris, Y., Pierrehumbert, R., {et~al.} 2006,
  Geochemistry, Geophysics, Geosystems, 7, doi:10.1029/2006GC001278

\bibitem[{Edson {et~al.}(2012)Edson, Kasting, Pollard, Lee, \&
  Bannon}]{Edson2012}
Edson, A.~R., Kasting, J.~F., Pollard, D., Lee, S., \& Bannon, P.~R. 2012,
  Astrobiology, 12, 562

\bibitem[{Fraedrich {et~al.}(2005)Fraedrich, Jansen, Kirk, Luksch, \&
  Lunkeit}]{Fraedrich2005}
Fraedrich, K., Jansen, H., Kirk, E., Luksch, U., \& Lunkeit, F. 2005,
  Meteorologische Zeitschrift, 14, 299

\bibitem[{{G{\'o}mez-Leal} {et~al.}(2016){G{\'o}mez-Leal}, {Codron}, \&
  {Selsis}}]{varysun}
{G{\'o}mez-Leal}, I., {Codron}, F., \& {Selsis}, F. 2016, \icarus, 269, 98

\bibitem[{Gow \& Williamson(1971)}]{volcanicalbedo}
Gow, A.~J., \& Williamson, T. 1971, Earth and Planetary Science Letters, 13,
  210

\bibitem[{{Greenberg} \& {Van Laerhoven}(2011)}]{christa}
{Greenberg}, R., \& {Van Laerhoven}, C. 2011, \apj, 733, 8

\bibitem[{{Haqq-Misra} {et~al.}(2016){Haqq-Misra}, {Kopparapu}, {Batalha},
  {Harman}, \& {Kasting}}]{haqqcycles}
{Haqq-Misra}, J., {Kopparapu}, R.~K., {Batalha}, N.~E., {Harman}, C.~E., \&
  {Kasting}, J.~F. 2016, \apj, 827, 120

\bibitem[{Hoffman {et~al.}(1998)Hoffman, Kauffman, Halverson, \&
  Schragg}]{hoffman98}
Hoffman, P., Kauffman, A., Halverson, G., \& Schragg, D. 1998, Science, 281,
  1342

\bibitem[{Hunter(2007)}]{Hunter:2007}
Hunter, J.~D. 2007, Computing In Science \& Engineering, 9, 90

\bibitem[{Kopparapu {et~al.}(2013)Kopparapu, Ramirez, Kasting,
  {et~al.}}]{kopparapu13}
Kopparapu, R., Ramirez, R., Kasting, J., {et~al.} 2013, The Astrophysical
  Journal, 765, 131

\bibitem[{Kump {et~al.}(2000)Kump, Brantley, \& Arthur}]{kump00}
Kump, L., Brantley, S., \& Arthur, M. 2000, Annual Review of Earth and
  Planetary Sciences, 28, 611

\bibitem[{Kyle {et~al.}(1981)Kyle, Jezek, Mosley-Thompson, \&
  Thompson}]{icevolcano2}
Kyle, P.~R., Jezek, P.~A., Mosley-Thompson, E., \& Thompson, L.~G. 1981,
  Journal of Volcanology and Geothermal Research, 11, 29

\bibitem[{{Le Hir} {et~al.}(2009){Le Hir}, Donnadieu, Godd{\'{e}}ris,
  Pierrehumbert, Halverson, Macouin, N{\'{e}}d{\'{e}}lec, \&
  Ramstein}]{Lehir2009}
{Le Hir}, G., Donnadieu, Y., Godd{\'{e}}ris, Y., {et~al.} 2009, Earth and
  Planetary Science Letters, 277, 453

\bibitem[{Lucarini {et~al.}(2010)Lucarini, Fraedrich, \&
  Lunkeit}]{lucarini2010}
Lucarini, V., Fraedrich, K., \& Lunkeit, F. 2010, Quarterly Journal of the
  Royal Meteorological Society, 136, 2

\bibitem[{Lucarini {et~al.}(2013)Lucarini, Pascale, Boschi, Kirk, \&
  Iro}]{Lucarini2013}
Lucarini, V., Pascale, S., Boschi, R., Kirk, E., \& Iro, N. 2013, Astronomische
  Nachrichten, 334, 576

\bibitem[{{Lunkeit} {et~al.}(2012){Lunkeit}, {Borth}, {B{\"o}ttinger},
  {et~al.}}]{plasimdocs}
{Lunkeit}, F., {Borth}, H., {B{\"o}ttinger}, M., {et~al.} 2012, Planet
  Simulator Reference Manual, 16th edn., National Center for Atmospheric
  Research

\bibitem[{Menou(2015)}]{menou2015}
Menou, K. 2015, Earth and Planetary Science Letters, 429, 20

\bibitem[{North {et~al.}(1981)North, Cahalan, \& {Coakley, Jr}}]{north1981}
North, G., Cahalan, R., \& {Coakley, Jr}, J. 1981, Reviews of Geophysics and
  Space Physics, 19, 91

\bibitem[{Pierrehumbert(2005)}]{pierrehumbert05}
Pierrehumbert, R. 2005, Journal of Geophysical Research, 110, 1

\bibitem[{Pierrehumbert(2010)}]{pierrehumbertbook}
---. 2010, Principles of Planetary Climate (New York: Cambridge University
  Press)

\bibitem[{{Poulsen} {et~al.}(2001){Poulsen}, {Pierrehumbert}, \&
  {Jacob}}]{dynamicocean}
{Poulsen}, C.~J., {Pierrehumbert}, R.~T., \& {Jacob}, R.~L. 2001, \grl, 28,
  1575

\bibitem[{Schwartzman \& Volk(1989)}]{schartz89}
Schwartzman, D., \& Volk, T. 1989, Nature, 340, 457

\bibitem[{Sternai {et~al.}(2016)Sternai, Caricchi, Castelltort, \&
  Champagnac}]{icevolcano1}
Sternai, P., Caricchi, L., Castelltort, S., \& Champagnac, J.-D. 2016,
  Geophysical Research Letters, 43, 1632, 2015GL067285

\bibitem[{{Tajika}(2007)}]{Tajika2007}
{Tajika}, E. 2007, Earth, Planets, and Space, 59, 293

\bibitem[{Turbet {et~al.}(2017)Turbet, Forget, Leconte, Charnay, \&
  Tobie}]{Turbet2017}
Turbet, M., Forget, F., Leconte, J., Charnay, B., \& Tobie, G. 2017, 1

\bibitem[{Voigt {et~al.}(2011)Voigt, Abbot, Pierrehumbert, \&
  Marotzke}]{Voigt2011}
Voigt, A., Abbot, D.~S., Pierrehumbert, R.~T., \& Marotzke, J. 2011, Climate of
  the Past, 7, 249

\bibitem[{Walker {et~al.}(1981)Walker, Hays, \& Kasting}]{walker81}
Walker, J., Hays, P., \& Kasting, J. 1981, Journal of Geophysical Research, 86,
  9776

\bibitem[{{Way} {et~al.}(2017){Way}, {Aleinov}, {Amundsen}, {Chandler},
  {Clune}, {Del Genio}, {Fujii}, {Kelley}, {Kiang}, {Sohl}, \&
  {Tsigaridis}}]{NASAGCM}
{Way}, M.~J., {Aleinov}, I., {Amundsen}, D.~S., {et~al.} 2017, ArXiv e-prints,
  arXiv:1701.02360

\bibitem[{{Williams} \& {Kasting}(1997)}]{wk97}
{Williams}, D.~M., \& {Kasting}, J.~F. 1997, \icarus, 129, 254

\end{thebibliography}

\end{document}